\documentclass[Physsubmission, Phys]{SciPost}
\hypersetup{
    colorlinks,
    linkcolor={red!50!black},
    citecolor={blue!50!black},
    urlcolor={blue!80!black}
}

\usepackage[bitstream-charter]{mathdesign}
\usepackage{amsmath}
\usepackage{bm}
\usepackage{empheq}
\usepackage{soul}
\newcommand{\mm}{\mathcal{M}}
\newcommand{\T}{{\bf T}}
\newcommand{\al}{\alpha}
\newcommand{\nn}{\nonumber}
\newcommand{\redeqn}{equation}
\newcommand{\beq}{\begin{\redeqn}}
\newcommand{\eeq}{\end{\redeqn}}
\newcommand{\beqa}{\begin{eqnarray}}
\newcommand{\eeqa}{\end{eqnarray}}

\newcommand{\eps}{\epsilon}
\newcommand{\tts}{\mathbf{T}_t^2}
\newcommand{\tsu}{\mathbf{T}_{s-u}^2}

\newcommand{\dd}{\mathbf{\Delta}}
\DeclareSymbolFont{usualmathcal}{OMS}{cmsy}{m}{n}
\DeclareSymbolFontAlphabet{\mathcal}{usualmathcal}
\DeclareMathAlphabet\mathbfcal{OMS}{cmsy}{b}{n}

\begin{document}

\begin{center}{\Large 
 \textbf{
The Soft Anomalous Dimension at four loops in the Regge Limit\\
}}\end{center}

\begin{center}
N. Maher \textsuperscript{1 $\star$}, G. Falcioni \textsuperscript{1}, E. Gardi \textsuperscript{1},  C. Milloy \textsuperscript{2},  L. Vernazza \textsuperscript{2,3} 
\end{center}

\begin{center}
{\bf 1} {Higgs Centre for Theoretical Physics, School of Physics and Astronomy,\\ The University of Edinburgh, Edinburgh EH9 3FD, Scotland, UK}
\\
{\bf 2} {Dipartimento di Fisica and Arnold-Regge Center, Universit\'{a} di Torino,\\
  and INFN, Sezione di Torino, Via P. Giuria 1, I-10125 Torino, Italy}
\\
{\bf 3} {Theoretical Physics Department, CERN, Geneva, Switzerland}
\\
* N.maher@sms.ed.ac.uk
\end{center}

\begin{center}
\today
\end{center}


\definecolor{palegray}{gray}{0.95}
\begin{center}
\colorbox{palegray}{
  \begin{tabular}{rr}
  \begin{minipage}{0.1\textwidth}
    \includegraphics[width=35mm]{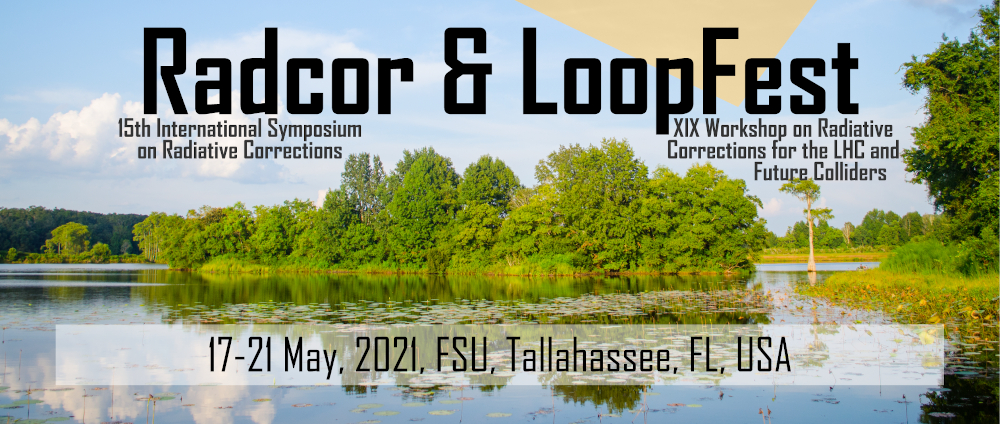}
  \end{minipage}
  &
  \begin{minipage}{0.85\textwidth}
    \begin{center}
    {\it 15th International Symposium on Radiative Corrections: \\Applications of Quantum Field Theory to Phenomenology,}\\
    {\it FSU, Tallahasse, FL, USA, 17-21 May 2021} \\
    \doi{10.21468/SciPostPhysProc.?}\\
    \end{center}
  \end{minipage}
\end{tabular}
}
\end{center}
\section*{Abstract}
{\bf
 The soft anomalous dimension governs the infrared divergences of scattering amplitudes in general kinematics to all orders in perturbation theory. By comparing the recent Regge-limit results for $2\to2$ scattering (through Next-to-Next-to-Leading Logarithms) in full colour to a general form for the soft anomalous dimension at four loops we derive
 powerful constraints on its kinematic dependence, opening the way for a bootstrap-based determination.
 }\\
\vspace{10pt}
\noindent\rule{\textwidth}{1pt}
\tableofcontents\thispagestyle{fancy}
\noindent\rule{\textwidth}{1pt}
\vspace{10pt}
\section{Introduction}
Scattering amplitudes are crucial to precision physics and our understanding of quantum field theory. Infrared singularities are a salient feature of gauge theory amplitudes, a manifestation of the gluon being massless.
 Infrared singularities factorise and exponentiate, and are thus governed by a finite and universal quantity -- the soft anomalous dimension \cite{Becher:2009cu,Becher:2009qa,Gardi:2009zv}. This quantity is therefore central to understand gauge theory scattering.
 The soft anomalous dimension for massless parton scattering in general kinematics is known to three-loop order \cite{Almelid:2015jia}. Furthermore its structure is highly constrained to all orders, owing to its intimate connection to the renormalisation of correlators of Wilson lines \cite{POLYAKOV:1980,Korchemsky:1985xj, Korchemsky:1987wg}.\\ 
 \indent
 The soft anomalous dimension for massless scattering can be defined as a correlator of a product of semi-infinite lightlike Wilson lines. This implies that its colour structure, at any loop order, corresponds to connected graphs and it admits bose symmetry under permutation of any two external lines, which links the kinematic dependence to that of colour. Finally, the kinematic dependence is largely constrained by the rescaling symmetry of individual Wilson line velocities.\\
 \indent The functional form of the three-loop soft anomalous dimension was recovered using a bootstrap approach in \cite{Almelid:2017qju}. In this approach, one writes down an ansatz for the kinematic functions using a suitable class of iterated integrals, and then fixes the (rational) coefficients in this ansatz using factorisation and symmetry constraints, along with constraints from the collinear and Regge limits. 
A general form in terms of colour factors multiplying unknown kinematic functions for the soft anomalous dimension for up to four loops was put forward in \cite{Becher:2019avh}.\\
 \indent The aim of the work reported in this talk is to find new  constraints on the unknown functions in the soft anomalous dimension at four loops \cite{Becher:2019avh} using information from a highly interesting kinematic limit: the Regge limit.  There has been much study of scattering amplitudes in the high-energy (Regge) limit in QCD \cite{Lipatov:1976zz,Balitsky:1978ic,Kuraev:1977fs,Collins:1977jy, Korchemskaya:1996je,DelDuca:2011ae}
including recent work on $2\to2$ scattering \cite{Grisaru:1973wbb, Grisaru:1973vw,Caron-Huot:2013fea,Caron-Huot:2017fxr, Caron-Huot:2017zfo,Gardi:2019pmk,Caron-Huot:2020grv, Vernazza:2021}. The new constraints are found by using results from newly calculated amplitudes in the Regge limit to Next-to-Leading Logarithmic (NLL) accuracy for the even signature amplitude ~\cite{Caron-Huot:2013fea,Lipatov:1976zz,Kuraev:1976ge,Caron-Huot:2017zfo} and NNLL accuracy for the odd signature amplitude \cite{Falcioni:2020lvv,Vernazza:2021,Reg3:2021}. These results are available in full colour for arbitrary representations of the scattered partons. They are compared to the soft anomalous dimension from \cite{Becher:2019avh} which is expressed in the Regge limit separated by signature. This work is discussed in detail in an upcoming paper \cite{Reg3:2021} with initial results published in \cite{Falcioni:2020lvv}. 

\subsection{Regge limit and signature}
\label{sec:intro}
Our focus is on $2\to2$ scattering shown diagrammatically in Fig. 1 below and expressed in the equation
\beqa\label{2to2}
i(p_1, a_1,\lambda_1) + j(p_2,a_2,\lambda_2) \to j(p_3,a_3,\lambda_3) + i(p_4,a_4,\lambda_4),
\eeqa
where the partons $i$, $j$ can each be a quark or a gluon and they respectively represent the target and projectile.
\begin{figure}[htp!]
 \centering
   \includegraphics[scale=0.7]{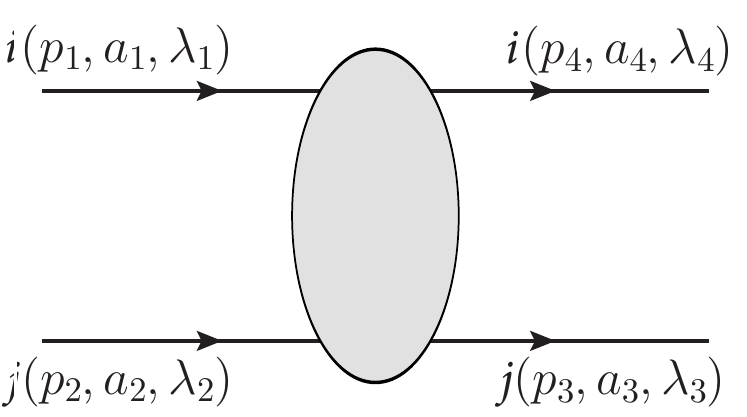}
\caption{2 to 2 particle scattering which at tree level correspond to eq.~\eqref{eq:treelevel}. The arrows indicate the direction of the momenta. The colour indices are $a_k$, while the labels $\lambda_k$ represent the particle helicities.}
\label{fig:2to2d}
\end{figure}Treating the particles with momenta $p_1$, $p_2$ as incoming, and $p_3$, $p_4$ as outgoing in Figure \ref{fig:2to2d}, the process is described in terms of the Mandelstam variables
\beq
s=(p_1+p_2)^2>0 \,\hspace{1cm}\,t=(p_1-p_4)^2<0\,\hspace{1cm}\,u=(p_1-p_3)^2<0.
\eeq
 For the tree-level diagram, the blob is a single-gluon t-channel exchange.
The tree-level expression is given by 
\beq
\label{eq:treelevel}
{\cal M}^{\rm tree}_{ij\to ij}= g_s^2\, \frac{2s}{t} \, \T_i \cdot \T_j \, \delta_{\lambda_1 \lambda_4} \delta_{\lambda_2 \lambda_3},
\eeq
where the factor $\delta_{\lambda_1 \lambda_4} \delta_{\lambda_2 \lambda_3}$ represents helicity conservation, and the colour dependence is expressed using the colour-space formalism introduced in \cite{Bassetto:1983mvz,Catani:1996jh,Catani:1996vz,Catani:1998bh}. Following this notation, a colour operator $\T_k$ corresponds to the colour generator associated with the $k$-th parton in the scattering amplitude, which acts as an SU($N_c$) matrix on the colour indices of that parton. In eq.~(\ref{eq:treelevel}), the colour indices $a_k$ of the incoming and outgoing partons are implicit. The dipole is expressed as  $\T_i \cdot \T_j = (T_i^b)_{a_1 a_4}(T_j^b)_{a_2 a_3}$, where $(T_i^b)_{a_1a_4} = t^b_{a_1 a_4}$ for quarks, $(T_i^b)_{a_1a_4} = - t^b_{a_4 a_1}$ for antiquarks, and $(T_i^b)_{a_1a_4} = i f^{a_1 b a_4}$ for gluons. \\
\indent
The high-energy limit is defined by the condition $ s \gg -t$,
i.e., the centre of mass energy $s$ becomes much larger than the momentum transfer $|t|$. As a result,
 $u \simeq -s$. Neglecting power-suppressed terms, this introduces an additional \emph{signature} symmetry to the amplitude under the exchange $s \leftrightarrow u$. It is then advantageous to split the amplitude into its even and odd components under $s\leftrightarrow u$: 
\begin{equation}
\label{eq:Msig}
 {\cal M}^{(\pm)}(s,t) = \tfrac12\Big( {\cal M}(s,t) \pm {\cal M}(-s-t,t) \Big), 
\end{equation}
where ${\cal M}^{(+)}$, ${\cal M}^{(-)}$
are referred to, respectively, as the \emph{even}
and \emph{odd} amplitudes. As demonstrated in ref.~\cite{Caron-Huot:2017fxr},
using the signature-even combination of logarithms,
\beq
\label{eq:siglogp}
L\equiv \log\left(\frac{s}{-t}\right)-\frac{i\pi}{2}=\log\left(\frac{-s- i0}{-t}\right)+\log\left(\frac{-u-i0 }{-t}\right),
\eeq
and expanding the amplitudes ${\cal M}_{\pm}$ according to 
\begin{equation}\label{eq:expansionDef}
{\mm}^{(\pm)}_{ij\to ij} = \sum_{n=0}^\infty \left(\frac{\al_s}{\pi}\right)^n \sum_{m=0}^nL^m{\mm}^{(\pm,n,m)}_{ij\to ij},
\end{equation}
with ${\mm}^{(-,0,0)}_{ij\to ij}\equiv {\cal M}_{ij\to ij}^{\rm tree}$, it can be shown that the odd amplitude coefficients $\mm^{(-,n,m)}_{ij\to ij}$ are purely \emph{real}, while the even ones $\mm^{(+,n,m)}_{ij\to ij}$ are purely \emph{imaginary}.
\subsection{Colour-operator notation}
Colour conservation in $2 \to 2$  scattering implies
\beq
\label{eq:colcons}
\left(\T_1+\T_2+\T_3+\T_4\right)\mathcal{M}_{ij\to ij}^{\rm{tree}}=0.\, 
\eeq 
In the high-energy limit it is helpful to express the colour generators using the basis of 
Casimirs corresponding to colour flow through the three channels~\cite{Dokshitzer:2005ig,DelDuca:2011ae}:
\beq\label{TtTsTu}
\T_s = \T_1+\T_2=-\T_3-\T_4 \hspace{0.9cm}\T_u = \T_1+\T_3=-\T_2-\T_4 \hspace{0.9cm}
\T_t  = \T_1+\T_4=-\T_2-\T_3\,.
\eeq
In order to make the signature symmetry manifest within Bose-symmetric amplitudes ${\cal M}^{(\pm)}$, it is useful to introduce a colour operator that is \emph{odd} under $s\leftrightarrow u$
crossing:
\beq
\label{eq:Tsudef}
\tsu\equiv \tfrac12 \left(\T_s^2-\T_u^2\right).
\eeq
This will form part of  what is called the \textit{Regge-limit basis}, which involves writing the colour operators in terms of $\tsu$ and $\tts$ in nested commutators where possible and can be seen in eqs.~(\ref{eq:softADdef4loopsymD6},\ref{eq:gammaNNLL+},\ref{eq:gammaNNLL-}).
The quartic Casimir  which contains a fully symmetrised trace, first appears in the soft anomalous dimension at four loops.
It is defined as
 \beq
 \label{eq:drri}
\frac{d_{R R_i}}{N_{R_i}}= \frac{1}{4}\sum_{\sigma\in\mathcal{S}_4}\mathrm{Tr}_R\left[T^{\sigma(a)} T^{\sigma(b)}T^{\sigma(c)}T^{\sigma(d)}\right]\mathbf{T}^a_i\mathbf{T}^b_i\mathbf{T}^c_i\mathbf{T}^d_i= \mathbfcal{D}^R_{iiii}
 \eeq
 where the $\mathbfcal{D}^R_{iiii}$ notation is adopted from~\cite{Becher:2019avh}.
\subsection{Soft anomalous dimension for $2\to2$ scattering}
The long-distance singularities of $2\to2$ massless amplitudes factorise following the infrared factorisation theorem for fixed-angle scattering 
~\cite{Catani:1998bh,Sterman:2002qn,Aybat:2006mz,Aybat:2006wq,Gardi:2009qi,Gardi:2009zv,Becher:2009cu,Becher:2019avh,Almelid:2015jia, Ma:2020}
\begin{align}
\label{ZH_intro}
    \mathcal{M}_{4}\left(s,t;\eps\right) =& \,\mathbf{Z}_{4}\left(s,t;\eps\right)
    \mathcal{H}_{4}\left(s,t;\eps\right)\,,
\end{align}
where $\epsilon$ is the dimensional regulator and the subscript is for $n=4$ partons. The infrared divergences are captured in $\mathbf{Z}_{4}$ which acts on the finite, so-called \emph{hard} function $\mathcal{H}_4$. The operator $\mathbf{Z}_{4}$ exponentiates
\begin{align}
        \mathbf{Z}_{4}\left(s,t;\eps\right) =&\, \mathcal{P}\exp{\left\{-\frac{1}{2}\int_0^{\mu^2}\frac{d\lambda^2}{\lambda^2}\bm{\Gamma}_4(s,t,\lambda,\al_s(\lambda^2))\right\}},\label{eq:ZfactorIntro}
\end{align}
where $\bm{\Gamma}_{4}$ is 
the soft anomalous dimension. The latter is a finite quantity depending on the $d$-dimensional running coupling $\alpha_s(\lambda^2)$. Its expansion is 
 \beq
\mathbf{\Gamma}_{4}(\{s_{ij}\},\lambda,\alpha_s)= \sum_{\ell=0}^{\infty} \left(\frac{\alpha_s}{\pi}\right)^{\ell}\mathbf{\Gamma}^{(\ell)}_{4}(\{s_{ij}\},\lambda),
 \eeq
where $\ell$ is the loop order. All kinematic functions within the soft anomalous dimension are expanded in a similar way. The soft anomalous dimension depends on the colour generators ${\bf T}_i$ associated to the external particles, which have been defined after eq.~\eqref{eq:treelevel}. The soft anomalous dimension also depends on the kinematic invariants
\beq
\label{eq:sij}
(-s_{ij})\equiv 2|p_i\cdot p_j| e^{-i \pi \sigma_{ij}},
\eeq 
where $p_i$ represents the momentum of the particle $i$, and $\sigma_{ij} = 1$ if both partons are in the initial or final state, otherwise $\sigma_{ij} = 0$. 
Specifically $\mathbf{\Gamma}_4$ depends on the invariants $s_{ij}$ either through the logarithms
\beq
\label{eq:lij}
l_{ij} \equiv \log\frac{-s_{ij}}{\lambda^2},
\eeq
 with $\lambda$ being the renormalisation scale, or via
  conformally invariant cross-ratios (CICRs)
\beq
\label{eq:beta}
\beta_{ijkl}=\log \rho_{ijkl}=\log\frac{(-s_{ij})(-s_{kl})}{(-s_{ik})(-s_{jl}) }=l_{ij}+l_{kl}-l_{ik}-l_{jl},
\eeq
 whose symmetries are discussed in \cite{Becher:2019avh,Dixon:2009ur}. It was shown in \cite{Becher:2009cu,Gardi:2009zv,Gardi:2009qi} that owing to soft-collinear factorisation and rescaling invariance of the Wilson-line velocities, the dependence on scale as in eq.~\eqref{eq:lij} is directly linked with collinear singularities and these are linear in $l_{ij}$ and are generated exclusively by the cusp anomalous dimension \cite{Becher:2009cu,Gardi:2009zv,Gardi:2009qi,Becher:2019avh}. The cusp anomalous dimension has the form \cite{Korchemsky:1987wg,Gardi:2009qi,Gardi:2009zv,Becher:2009cu,Becher:2019avh}
 \beq
\label{cuspAD}
\Gamma^{\rm{cusp}}_i(\alpha_s(\lambda^2)) = \frac{1}{2} \gamma_K(\alpha_s(\lambda^2))C_i+ \sum_R g_R(\alpha_s(\lambda^2)) \frac{d_{RR_i}}{N_i} + {\cal}O(\alpha_s^5),
\eeq
where $\gamma_K$ multiplies the quadratic Casimir $C_i$ in the representation of parton~$i$, while the component $g_R$, starting only at four loops, multiplies the quartic Casimir (defined in eq.~(\ref{eq:drri})). $\Gamma_i^{\rm{cusp}}$ is known to four loops in QCD \cite{Boels:2017ftb,Boels:2017skl,Moch:2017uml,Grozin:2017css,Henn:2019swt}.
  In contrast, dependence on the kinematics through conformal cross ratios in eq.~\eqref{eq:beta} is not constrained by factorisation and can be complicated. It has been computed at three loops in refs.~\cite{Almelid:2015jia,Almelid:2017qju} and is yet unknown at four loops.\\
 \indent The soft anomalous dimension for $2\rightarrow 2$ scattering through four loops may be expressed as \cite{Becher:2019avh}
 \begin{align}\begin{split}
 \label{eq:SADintrogammap}
 \mathbf{\Gamma}_4(\{s_{ij}\},\lambda, \alpha_s(\lambda^2)) &=-\frac{1}{4} \gamma_{K}(\alpha_s)\sum_{(i,j)}\,\T_i\cdot\T_j\,\log\left(\frac{-s_{ij}}{\lambda^2}\right) +\sum_i^4\,\gamma_i(\alpha_s)
 \\
   &\quad\mbox{} + f(\alpha_s) \sum_{(i,j,k)}\mathbfcal{T}_{ijki}   +\!\sum_{(i,j,k,l)} \mathbfcal{T}_{ijkl}\,{\cal F}(\beta_{ijlk},\beta_{iklj};\alpha_s)  \\
   &\quad\mbox{}-\sum_R \frac{g_R(\alpha_s)}{2} \bigg[ \sum_{(i,j)}\,\big( \mathbfcal{D}_{iijj}^R + 2 \mathbfcal{D}_{iiij}^R \big)  \log\left(\frac{-s_{ij}}{\lambda^2}\right) + \sum_{(i,j,k)} \mathbfcal{D}_{ijkk}^R\,\log\left(\frac{-s_{ij}}{\lambda^2}\right)\bigg] \\
   &\quad\mbox{}
    + \sum_R \sum_{(i,j,k,l)}\!\mathbfcal{D}_{ijkl}^R\,{\cal G}_R(\beta_{ijlk},\beta_{iklj};\alpha_s)+ \sum_{(i,j,k,l)}\!\mathbfcal{T}_{ijkli}\,{\cal H}_1(\beta_{ijlk},\beta_{iklj};\alpha_s)  \,,
     \end{split}\end{align}
  the summation is over permutations of lines $(1,2,3,4)$. The various terms in eq.~\eqref{eq:SADintrogammap} correspond to colour connected structures as dictated by the non-Abelian exponentiaion theorem \cite{Gatheral1983ExponentiationOE,Frenkel1984NonabelianEE,Gardi:2013ita}. The colour structures have been defined in \cite{Becher:2019avh} as
  \begin{equation}\label{symstruc}
\begin{aligned}
    \mathbfcal{T}_{ijkl} &=\frac{1}{4!} f^{ade} f^{bce} \,\sum_{\sigma\in S_4}\,\mathbf{T}_i^{\sigma(a)}\mathbf{T}_j^{\sigma(b)},\mathbf{T}_k^{\sigma(c)}\mathbf{T}_l^{\sigma(d)},  \\
   \mathbfcal{D}_{ijkl}^R 
   &= \frac{1}{4!}\sum_{\sigma \in S_4} \operatorname{Tr}_R\left(T^{\sigma(a)}T^{\sigma(b)}T^{\sigma(c)}T^{\sigma(d)}\right)\,\T_i^a \T_j^b \T_k^c \T_l^d \,,
    \\
   \mathbfcal{T}_{ijklm} 
   &= \frac{1}{5!}i f^{adf}f^{bcg}f^{efg} \sum_{\sigma\in S_5}\,\mathbf{T}_i^{\sigma(a)}\mathbf{T}_j^{\sigma(b)}\mathbf{T}_k^{\sigma(c)}\mathbf{T}_l^{\sigma(d)}\mathbf{T}_m^{\sigma(e)}.
\end{aligned}
\end{equation}
There is an explicit sum over representations $R$ of the particle content of the specific theory considered e.g QCD, in the functions $g_R$ and ${\cal G}_R$ in the third and fourth lines.
The coupling depends on the scale $\al_s(\lambda^2)$ throughout but we will just write $\alpha_s$ from here onwards. 
\\
\indent At one and two loop order, only the first line of eq.~\eqref{eq:SADintrogammap}, the so-called dipole formula contributes. The collinear anomalous dimension for parton $i$ is $\gamma_i$, studied in refs. ~\cite{FormFactors,DelDuca:2014cya,Falcioni:2019nxk,Dixon:2017nat}. It has recently been computed at four loops in QCD~\cite{Agarwal:2021zft}. 
  The terms on the second line first appear at three loops with their known expressions from \cite{Almelid:2015jia}. At four-loop order there is an implicit sum over the representations contributing to the $f$ and ${\cal F}$ colour structures. The terms from the third line onwards in eq.~\eqref{eq:SADintrogammap} appear for the first time at four loops. The fully-symmetric function ${\cal G}_R(\beta_{ijlk},\beta_{iklj};\alpha_s)$ depends on CICRs and the representation. The function ${\cal H}_1$ depends on CICRs and does not depend on the representation at four-loop order with an expected sum appearing at five loops and beyond. Our focus will be on understanding the high-energy limit of the unknown functions that depend on CICRs at four loops.
\section{Soft Anomalous Dimension in the  high-energy limit}
In the high-energy limit of $2\to 2$ scattering, the soft anomalous dimension $\mathbf{\Gamma}_4$ takes the form~\cite{DelDuca:2011wkl,Caron-Huot:2017fxr,DelDuca:2011ae}
\begin{align}
 \begin{split}
\label{eq:softADdef6p}
\mathbf{\Gamma}_{4}\left(\alpha_s(\lambda^2),L,\frac{-t}{\lambda^2}\right)=&\,\frac{1}{2}\gamma_{K}(\alpha_s)\left[L\tts + i\pi\tsu\right] +   \Gamma_i\left(\alpha_s,\frac{-t}{\lambda^2}\right)+\Gamma_j\left(\alpha_s,\frac{-t}{\lambda^2}\right)\\&+\sum_{\ell=3}^\infty \left(\frac{\alpha_s}{\pi}\right)^\ell \sum^{\ell-1}_{m=0} L^m \dd^{(\ell,m)},
\end{split}
\end{align} 
 where $L$ is the signature-even log of eq.~\eqref{eq:siglogp}. $\Gamma_i$ captures the collinear singularities for parton $i$ and contains the collinear anomalous dimension $\gamma_i$ and the cusp anomalous dimension eq.~\eqref{cuspAD}. The dipole formula in the  first line of eq.~(\ref{eq:softADdef6p}) is exact up to two loops~\cite{Gardi:2009qi,Becher:2009cu}. All terms that are not part of the dipole formula are collected in the second line. The known coefficients $\dd^{(\ell,m)}$ are given in Appendix \ref{sec:SADhej}.\\
 \indent In the high-energy limit, all the terms in the anomalous dimension of eq.~\eqref{eq:softADdef6p} are expanded in powers of the large logarithm $L$. Furthermore, we separate terms by their signature under crossing $s \leftrightarrow u$. The contributions proportional to $\tts, \Gamma_i,\Gamma_j$ in the first line are manifestly even under $s\leftrightarrow u$, while $\tsu$ is odd by definition. The coefficients $\dd^{(\ell,m)}$ decompose according to their signature in $\dd^{(\pm,\ell,m)}$, similarly to the complete amplitude in eq.~\eqref{eq:Msig}. All the coefficients in eq.~\eqref{eq:softADdef6p} at three loops are known by expanding the result in general kinematics \cite{Caron-Huot:2017fxr}.  It is worth recalling that, upon expansion, the soft anomalous dimension will be multiplied by the odd tree-level amplitude in eq.~\eqref{eq:treelevel}: for this reason, odd signature in the amplitude corresponds to even signature in the soft anomalous dimension.
 At four loops, the explicit calculations of the NLL in the signature-even amplitude \cite{Caron-Huot:2013fea,Caron-Huot:2017zfo, Gardi:2019pmk, Caron-Huot:2020grv} and of the NNLL in the signature-odd one \cite{Falcioni:2020lvv,Reg3:2021} yield 
 \begin{align}
 \begin{split}
\label{eq:softADdef4loopsymD6}
\mathbf{\Gamma}^{(4)}_4(L) = &- L^3 i \pi \frac{\zeta_3}{24}\Big[\tts,[\tts,\tsu]\Big]\tts+L^2\dd^{(-,4,2)}\\&+L^2 \zeta_2 \zeta_3 \bigg(\frac{d_{AA}}{N_A} -\frac{C_A^4}{24}-\frac{1}{4}\tts[(\tsu)^2,\tts]+\frac{3}{4}[\tsu,\tts]\tts\tsu\bigg)\\& + {\cal O}(L),
\end{split}
\end{align}
where $\dd^{(-,4,2)}$  and corrections at $O(L)$ and $O(L^0)$ are still to be determined. Despite this result being valid for QCD, there are no $n_f$ terms present in eq.~\eqref{eq:softADdef4loopsymD6}. This is so because the relevant contributions (at this logarithmic accuracy) are generated solely by gluon diagrams. Below  we use the result in eq.~\eqref{eq:softADdef4loopsymD6} to constrain the unknown functions in eq.~\eqref{eq:SADintrogammap}.
 \subsection{Separating the soft anomalous dimension by signature and colour operators}\label{sec:factor2to2}
  In the previous section, at four-loop order, we have presented a newly calculated expression for the soft anomalous dimension in high-energy limit to NNLL which contains only colour adjoint contributions.
  In comparing eq.~\eqref{eq:softADdef4loopsymD6} with eq.~\eqref{eq:SADintrogammap}  it becomes clear that any sums over the representations $R$ collapse to the adjoint representation and the expression becomes
                 \begin{align}\begin{split}
 \label{eq:SADNNLL}
 \mathbf{\Gamma}^{(4)}_4(\{\beta_{ijkl}\}) &= 
   \!\sum_{(i,j,k,l) \in S_4}\!\mathbfcal{T}_{ijkl}\,{\cal F}_A^{(4)}(\beta_{ijlk},\beta_{iklj})
    +  \sum_{(i,j,k,l) \in S_4}\!\mathbfcal{D}^A_{ijkl}\,{\cal G}_A^{(4)}(\beta_{ijlk},\beta_{iklj})\\
   &\quad\mbox{}+ \sum_{(i,j,k,l) \in S_4}\!\mathbfcal{T}_{ijkli}\,{\cal H}_1^{(4)}(\beta_{ijlk},\beta_{iklj})+ \mathcal{O}(L).
        \end{split}\end{align}
The superscript is for the loop order $\ell=4$. The four-loop order coefficients $(\gamma_{K,R}^{(4)},g_R^{(4)},f^{(4)}_R$) contributing at $O(L)$ and $O(L^0)$ are suppressed and are discussed in ref.~\cite{Reg3:2021}.\\
  \indent
 Taking the Regge limit of the kinematic functions involves an analytical continuation to the physical region, and an expansion of the functions in powers of the high-energy signature-even logarithm $L$. This procedure has been discussed in detail for the three- loop case in~\cite{Almelid:2017qju}. Here we consider the four-loop case, where an explicit calculation in general kinematics is still missing. 
 In the Regge limit, an expansion in the signature-even logarithm $L$ can be performed on each of the kinematic functions. For example
\beq
{\cal F}^{(+,4)}_A(L)={\cal F}^{(+,4,3)}_A L^3 + {\cal F}^{(+,4,2)}_A L^2+{\cal F}^{(+,4,1)}_A L +{\cal F}^{(+,4,0)}_A,
\eeq
with all other functions contributing at NNLL accuracy in the Regge limit having a similar expansion.
  The colour structures are expressed in a \textit{Regge-limit basis} with the steps elaborated in \cite{Reg3:2021}. The subscript $\text{Regge}$ denotes functions after the Regge limit has been taken.
The signature-even part is
      \beq
\label{eq:gammaNNLL+}\begin{aligned}
\mathbf{\Gamma}^{(+,4)}_{4,\text{Regge}}=&2{\cal F}_{A}^{(+,4)} (L)\Big[\tsu, [\tsu,\tts]\Big]  \\
      &\quad\mbox{}
    +  {\cal G}_{A}^{(+,4)}(L)\bigg(2\left(\frac{d_{AA}}{N_A} -\frac{C_A^4}{24}\right)  -\frac{1}{2}\tts[(\tsu)^2,\tts]+\frac{3}{2}[\tsu,\tts]\tts\tsu \\
      &\quad\mbox{}+\frac{C_A}{2} \Big[\tsu,[\tsu,\tts]\Big] 
    \bigg) +{\cal H}_{1}^{(+,4)}(L)\bigg(-\frac{1}{2}C_A\Big[\tsu,[\tsu,\tts]\Big]\\&\quad\mbox{}+\frac{3}{4}C_A \tsu [\tsu,\tts] -\frac{1}{4}\tts[\tsu,\tts]\tsu\bigg)+ {\cal O}(L),
      \end{aligned}
      \eeq
            and the signature-odd part 
            \beq
\label{eq:gammaNNLL-}\begin{aligned}
\mathbf{\Gamma}^{(-,4)}_{4,\text{Regge}}&=-{\cal F}_{A}^{(-,4)}(L) \Big[\tts, [\tts,\tsu]\Big] \\ &\quad\mbox{}
      +{\cal H}_{1}^{(-,4)}(L)\bigg(-\frac{1}{2}\Big[\tsu,[\tsu,[\tsu,\tts]\Big]+\frac{1}{8}\Big[\tts,\big[\tts,[\tts,\tsu]\big]\Big]\bigg)
      \\ &\quad\mbox{}+\tilde{{\cal H}}_{1}^{(-,4)}(L)\left(\frac{1}{4}\Big[\tts,\big[\tts,[\tts,\tsu]\big]\Big]\right)+ {\cal O}(L).
          \end{aligned}
          \eeq
The definitions of the even and odd functions of ${\cal F}^{(\pm)}$ and ${\cal H}_1^{(\pm)}$ are given in Appendix \ref{EOA}. ${\cal G}_A$ is a completely symmetric function so ${\cal G}_A^{(-)}=0$ and it only contributes to the signature-even part of the soft anomalous dimension. are given in Appendix \ref{EOA}. The quartic Casimir of $\frac{d_{AA}}{N_A}$ is defined in eq.~\eqref{eq:drri}. Both eqs.~(\ref{eq:gammaNNLL+},\ref{eq:gammaNNLL-}) are fully non-planar. The planar contributions appear at ${\cal O}(L^1)$ and are associated with the cusp anomalous dimension with more discussion in \cite{Reg3:2021}.
\section{Constraint Results}
 We now compare eqs.~(\ref{eq:gammaNNLL+},\ref{eq:gammaNNLL-}), which correspond to the most general form for the soft anomalous dimension at four-loop order, expanded in the Regge limit through quadratic terms in $L$, with the results of explicit calculations of the soft anomalous dimension in eq.~\eqref{eq:softADdef4loopsymD6}.
  We obtain the  set of constraints for the unknown kinematic functions summarised in Table. \ref{tab:reggeconstraints}. 
\begin{table}[h]
\begin{center}
\begin{tabular}{ ||c|c|c||c|c|c|| } 
 \hline
$+$& $L^3$&$L^2$&$-$ & $L^3$&$L^2$\\
\hline
  ${\cal F}^{(+,4)}_A$ & 0 & $-C_A\frac{\zeta_2\zeta_3}{8}$&${\cal F}^{(-,4)}_A$ & $ i \pi C_A \frac{\zeta_3}{24}$& ? \\ 
   ${\cal F}^{(+,4)}_F$ & 0 & 0 & ${\cal F}^{(-,4)}_F$ & 0 & ?\\ 
   $  {\cal G}^{(+,4)}_A$ & 0 & $\frac{\zeta_2\zeta_3}{2}$ & ${\cal G}^{(-,4)}_R$ & 0 & 0\\       $  {\cal G}^{(+,4)}_F$ & 0 & 0 &${\cal H}^{(-,4)}_1$ & 0 & ? \\ 
   ${\cal H}^{(+,4)}_1$ & 0 &0& $\tilde{{\cal H}}^{(-,4)}_1$ & 0 & ? \\  
   \hline
\end{tabular}
 \caption{Regge constraints separated by signature for the kinematic functions at four-loop order for $2 \to 2$ scattering  in soft anomalous dimension ansatz of eq.~\eqref{eq:SADintrogammap}. } \label{tab:reggeconstraints}
 \end{center}
\end{table}
 In the left columns of the table, we show the signature-even functions to NNLL accuracy. The signature-odd functions are given in the right columns of the table with constraints to NLL accuracy. The question marks represent quantities which have not been calculated. Because the even part of the soft anomalous dimension is two-loop exact at NLL and since the functions in eq.~\eqref{eq:gammaNNLL+} all multiply independent colour structures, they are each individually zero at $O(L^3)$.  Since ${\cal G}_R$ is a purely symmetric function, any antisymmetric component would be zero at all orders of $L$.
 We notice that only two functions contribute to $\Gamma^{(+,4)}$ through the terms proportional to $L^2$.
\section{Conclusion}
Using the state-of-the art knowledge of $2\to2$ scattering amplitudes in the high-energy limit we determined NLL and NNLL contributions to the soft anomalous dimension at four loops.
These provide constraints on the unknown functions in the soft anomalous dimension. We derive new inhomogeneous constraints to the general form of the soft anomalous dimension in general kinematics in \cite{Becher:2019avh}, reported in Table \ref{tab:reggeconstraints}.
\\
\indent Our results are consistent with the work of \cite{Vladimirov:2017ksc}, which argues that the function ${\cal H}_1$ must vanish exactly on the basis of symmetry considerations which exclude terms with an odd number of generators in the soft anomalous dimension.
 These constraints from the Regge limit along with those from the collinear limit in \cite{Becher:2019avh} are important input for a bootstrap approach to determine the unknown functions in the soft anomalous dimension in general kinematics at four loops with a similar method  to \cite{Almelid:2017qju}. 
\section*{Acknowledgements}
We would like to thank Simon Caron-Huot for insightful comments and Claude Duhr and Andrew McLeod for collaboration on a related project on the soft anomalous dimension. 
\paragraph{Funding information}
EG, GF and NM are supported by the STFC Consolidated Grant ‘Particle Physics at the Higgs Centre’. GF is supported by the ERC Starting Grant 715049 ‘QCDforfuture’ with Prinipal Investigator Jennifer Smillie. CM's work is supported by the Italian Ministry of University and Research (MIUR), grant PRIN 20172LNEEZ. LV is supported by Fellini, Fellowship for Innovation at INFN, funded by the 
European Union's Horizon 2020 research programme under the Marie Sk\l{}odowska-Curie Cofund Action, grant agreement no. 754496.
\begin{appendix}
\section{Even and odd functions in the soft anomalous dimension}
\label{EOA}
The even and odd functions under $s\leftrightarrow u\,(2 \leftrightarrow 3)$ are
  \begin{align}
        \label{eq:Fsigeven}
        {\cal F}^{(+)}(\{\beta_{ijkl}\},\alpha_s)&\equiv  \frac{1}{2}\bigg\{{\cal F}(\beta_{1324},\beta_{1423};\alpha_s
)+ {\cal F}(\beta_{1234},\beta_{1432};\alpha_s)\bigg\},\\
{\cal F}^{(-)}(\{\beta_{ijkl}\},\alpha_s)&\equiv  \frac{1}{2}\bigg\{{\cal F}(\beta_{1234},\beta_{1432};\alpha_s)-{\cal F}(\beta_{1324},\beta_{1423};\alpha_s
)\bigg\}+{\cal F}(\beta_{1243},\beta_{1342};\alpha_s).\\
\nn\\
        \label{eq:Hsig}
   {\cal H}_1^{(+)}(\{\beta_{ijkl}\}, \alpha_s)&\equiv\frac{1}{2}\Big\{ {\cal H}_1(\beta_{1324},\beta_{1423};\alpha_s)+ {\cal H}_1(\beta_{1234},\beta_{1432};\alpha_s)\Big\},\\
{\cal H}_1^{(-)}(\{\beta_{ijkl}\}, \alpha_s)&\equiv\frac{1}{2}\Big\{ {\cal H}_1(\beta_{1324},\beta_{1423};\alpha_s)- {\cal H}_1(\beta_{1234},\beta_{1432};\alpha_s)\Big\},
\\
\tilde{{\cal H}}_1^{(-)}(\{\beta_{ijkl}\}, \alpha_s)&\equiv {\cal H}_1(\beta_{1243},\beta_{1342};\alpha_s).
    \end{align}
    These are required in order to separate the soft anomalous dimension by signature in the high-energy limit and appear in eqs. (\ref{eq:gammaNNLL+},\ref{eq:gammaNNLL-}).
    \section{Soft Anomalous dimension in the high-energy limit from three loops}
    \label{sec:SADhej}
    The soft anomalous dimension at one and two loops is captured by the dipole formula~\cite{Gardi:2009zv,Becher:2009cu}
\beq\label{eq:DipoleAppendix}
\mathbf{\Gamma}_n^{\rm dip.}(\{s_{ij}\},\lambda,\alpha_s)=-\frac{\gamma_{K}(\alpha_s)}{4}\, \sum_{(i,j)}\,\T_i\cdot\T_j\,\,l_{ij} +\sum_i^n\,\gamma_i(\alpha_s),
\eeq
with the $l_{ij}=\log \frac{-s_{ij}}{\lambda^2}$. The anomalous dimension of parton  $i$ is $\gamma_i(\alpha_s)$, one for each external particle $i$. 
The soft anomalous dimension in the Regge limit for $2\to 2$ scattering is given in eq.~\eqref{eq:softADdef6p} and at three-loop order, it reads
\beq
\mathbf{\Gamma}^{(3)}_{ij\to ij}\left(L,\frac{-t}{\lambda^2}\right) = \sum_R\frac{\gamma_{K,R}^{(3)}}{2}\left[L\tts + i\pi\tsu\right] +\Gamma_i^{(3)}\left(\frac{-t}{\lambda^2}\right)+\Gamma_j^{(3)}\left(\frac{-t}{\lambda^2}\right) +\sum_{m=0}^{2}\dd^{(3,m)} L^m
\eeq
with $\dd^{(3,m)}$ being corrections to the dipole formula starting at three loops. The corrections below were calculated explicitly in \cite{Almelid:2015jia}, after which colour expressions with signature in the Regge limit were found in \cite{Caron-Huot:2017zfo}
     \begin{align}
     \label{eq:delta32+}
     \dd^{(-,3,2)}&=\text{Im}\left[\dd^{(3,2)}\right]=0\\
     \label{eq:delta32-}
     \dd^{(+,3,2)}&=\text{Re}\left[\dd^{(3,2)}\right]=0\\
     \dd^{(-,3,1)}&=i \pi \Big[\tts,[\tts,\tsu]\Big] \frac{11}{4} \zeta_3\\
           \label{eq:delta31}
     \dd^{(+,3,1)}&=\text{Re}\left[\dd^{(3,1)}\right]=0\\
           \label{eq:delta30plus}
     \dd^{(-,3,0)}&=i \pi \Big[\tts,[\tts,\tsu]\Big] \frac{11}{4} \zeta_4
     \\
      \label{eq:delta30min}
     \dd^{(+,3,0)}&=\frac{1}{4} \Big[ \tsu, [\tsu,\tts]\Big]\Big(4\zeta_2 \zeta_3-\zeta_5\Big)- \frac{\zeta_5 +2 \zeta_2 \zeta_3}{8}\bigg\{ - \frac{5}{8} C_A^2 \tts+ f^{abe}f^{cde} \nn\\\quad\mbox{}& \times \Big[ \{\T_t^a, \T_t^d\}\left(\{\T_{s-u}^b, \T_{s-u}^c\}+\{\T_{s+u}^b, \T_{s+u}^c\}\right)+ \{\T_{s-u}^a,\T_{s-u}^d\}\{\T^b_{s+u}, \T_{s+u}^c\}\Big] \bigg\}.
     \end{align}

         The generators in the line above are defined  as 
     \begin{align}
     \T_{s-u}^a\equiv \frac{1}{\sqrt{2}}\left(\T_s^a-\T_u^a\right)\hspace{1cm}\T_{s+u}^a\equiv \frac{1}{\sqrt{2}}\left(\T_s^a+\T_u^a\right).
         \end{align}
     At four-loop order, the soft anomalous dimension in the high-energy limit is
\beq
\mathbf{\Gamma}^{(4)}_{ij\to ij}\left(L,\frac{-t}{\lambda^2}\right) = \sum_R\frac{\gamma_{K,R}^{(4)}}{2}\left[L\tts + i\pi\tsu\right] +\Gamma_i^{(4)}\left(\frac{-t}{\lambda^2}\right)+\Gamma_j^{(4)}\left(\frac{-t}{\lambda^2}\right) +\sum_{m=0}^{3}\dd^{(4,m)} L^m.
\eeq
     At four loops, the specific corrections are 
      \begin{align}
     \label{eq:delta43+}
     \dd^{(+,4,3)}&=\text{Re}\left[\dd^{(3,2)}\right]=0,\\
     \label{eq:delta43-}
     \dd^{(-,4,3)}&=-i\pi\frac{\zeta_3}{24} \Big[\tts,[\tts,\tsu]\Big]\tts,\\
     \dd^{(+,4,2)}&=\zeta_2\zeta_3\bigg\{-\frac{C_A^4}{24}+\frac{d_{AA}}{N_A} + \frac{1}{4}(-\tts[(\tsu)^2,\tts]+3[\tsu,\tts]\tts\tsu)\bigg\},
           \label{eq:delta42+}
     \end{align}
     where $\dd^{(-,4,2)}$  is still to be found and corrections at $O(L)$ and $O(L^0)$ still to be determined.
     \end{appendix}
\bibliography{mybib.bib}
\nolinenumbers
\end{document}